\documentclass[prb,twocolumn,showpacs,floatfix]{revtex4}
\usepackage{graphicx}
\usepackage{bm}
\usepackage{dcolumn}
\begin{document}

\title {A model for the onset of transport in systems with 
distributed thresholds for conduction}

\author{Klara Elteto, Eduard G. Antonyan, T. T. Nguyen, and 
Heinrich M. Jaeger}
\affiliation{James Franck Institute and Department of Physics,
   University of Chicago, Chicago, IL 60637}

\date{\today}

\begin{abstract}
We present a model supported by simulation to explain the effect of 
temperature on the conduction threshold in disordered systems. Arrays 
with randomly distributed local thresholds for conduction occur in 
systems ranging from superconductors to metal nanocrystal arrays. 
Thermal fluctuations provide the energy to overcome some of the local 
thresholds, effectively erasing them as far as the global conduction 
threshold for the array is concerned. We augment this thermal energy 
reasoning with percolation theory to predict the temperature at which 
the global threshold reaches zero.
We also study the effect of capacitive nearest-neighbor interactions on the 
effective charging energy. Finally, we present results from Monte 
Carlo simulations that find the lowest-cost path across an array as a 
function of temperature. The main result of the paper is the linear 
decrease of conduction threshold with increasing temperature: $V_t(T) 
= V_t(0) (1 - 4.8 k_BT P(0)/ p_c ) $, where $1/P(0)$ is an effective 
charging energy that depends on the particle radius and interparticle 
distance, and $p_c$ is the percolation threshold of the underlying lattice. 
The predictions of this theory compare well to experiments in one- 
and two-dimensional systems.
\end {abstract}


\maketitle

\section{Introduction}
In many physical systems, local barriers prevent the onset of 
steady-state motion or conduction unless a certain minimum threshold 
for an externally applied driving force or bias is exceeded.   Often, 
the strength of those barriers varies throughout the system and only 
their statistical distribution is known.  A key issue then concerns 
how the global threshold for onset of motion is related to the 
distribution of local threshold values.  Examples include the onset 
of resistance due to depinning of fluxline motion in type-II 
superconductors, the onset of mechanical motion in coupled frictional 
systems such as sand piles, and the onset of current flow through 
networks of tunnel junctions in the Coulomb blockade regime.  In all 
of these cases, defects in the host material or the underlying 
substrate produce local traps or barriers of varying strength.

Under an applied driving force, fluxlines, mobile particles or charge 
carriers from an external reservoir can penetrate the disordered 
energy landscape, becoming stuck at the traps or piling up in front 
of barriers.  With increased drive, particles can surmount some of 
the barriers and penetrate further. However, a steady-state flow is 
only established once there is at least one continuous path 
connecting one side of the system with the other.  The onset of 
steady-state transport then corresponds to finding the lowest-energy 
system-spanning path. This optimization problem was addressed in 1993 
in a seminal paper by Middleton and Wingren (MW).\cite{MW}

Using analytical arguments as well as computer simulations, MW found 
that, for the limit of negligible thermal energies, the onset of 
system-spanning motion corresponds to a second order phase transition 
as a function of applied bias.  The global threshold value scales 
with distance across the system, but is independent of the details of 
the barrier size distribution.  Beyond threshold, more paths open up 
and the overall transport current increases. As a result, the 
steady-state transport current displays power law scaling as a 
function of excess bias.  These predictions have subsequently been 
used extensively in the interpretation of single electron tunneling 
data from networks of lithographically defined junction arrays\cite{rimberg,kurdak} as well as from self-assembled nanoparticle 
systems.\cite{wybourne,ancona,Black}  In addition, recent experiments\cite{parthasarathy01} and simulations\cite{reichhardt03a} have 
explored how the power law scaling is affected by structural disorder 
in the arrays. The regime of large structural disorder and 
significant voids in the array was investigated numerically using a 
percolation model.\cite{rendell}

What happens at finite temperature? Intuitively, one might expect 
temperature to produce a smearing of the local thresholds and thus a 
quick demise of the power law scaling for $T > 0$.  Indeed, a number 
of experiments have found that the nonlinear current-voltage 
characteristics observed at the lowest temperatures give way to 
nearly linear, Ohmic behavior once $T$ is raised to a few dozen 
Kelvin.\cite{beverly,sampaio}  More recently, however, several 
experiments showed that the scaling behavior survives with a 
well-defined, albeit now temperature-dependent, global threshold.  In 
a previous Letter, we demonstrated for a 
two-dimensional metal nanocrystal array that a) the threshold is only 
weakly temperature dependent, decreasing linearly with increasing 
$T$, and b) the scaling exponent remains unaffected by temperature. 
Consequently, the shape of the nonlinear response as a function of 
applied drive remains constant and is merely shifted to lower drive 
values as $T$ increased.\cite{parthasarathy04}

Similar behavior was also observed in small 2D metal nanoparticle 
networks by Ancona {\em et al.}\cite{ancona} and Cordan {\em et al.}\cite{cordan} and in 1D chains of 
carbon particles by Bezryadin. {\em et al.}\cite{bezryadin}  Most 
recently, it was corroborated by simulations of (semi-classical) 
particles in 2D arrays of pinning sites with random strengths.\cite{reichhardt03b} This weak temperature dependence of the 
nonlinear response also has important practical consequences as it 
implies that arrays are much more robust and forgiving as compared to 
systems with a single threshold that might be significantly affected 
by its local environment.

However, the theoretical approach developed by MW considers only the 
zero-temperature limit where the local energy levels are sharply 
delineated and barriers between adjacent sites are well-defined.  In 
Ref. \onlinecite{parthasarathy04} we introduced the main results from a new 
model that extends the MW approach to finite temperatures.  Here, we 
develop this model in more detail, providing both analytical results 
and data from computer simulations. For concreteness, we focus on 
single electron tunneling through metal nanoparticle arrays. 
However, we expect the main results to carry over to a much wider 
class of systems with distributed thresholds due to quenched disorder.

Our model goes beyond previous work in two important aspects. First, 
we introduce a method that allows us to treat finite temperatures. 
This method is based on estimating when small barriers, washed out by 
temperature, have percolated across the system, and it establishes an 
upper limit on the global threshold as a function of temperature.  A 
key finding is that random quenched disorder leads to universal 
behavior that is independent of the details of the barrier height 
distribution. Second, we include nearest neighbor capacitive 
coupling.  This leads us to a new definition of the relevant 
effective charging energy for system crossing, in terms of the most 
probable value in the distribution of energy costs. As shown in Ref. \onlinecite{parthasarathy04}, the model captures the experimentally 
observed temperature dependence of the drive-response characteristics 
and predicts the collapse of the global threshold as a function of 
temperature on a universal curve that is independent of local 
junction parameters.

The paper is organized as follows.  In Section II we outline the 
basic ingredients of our model, mainly focusing on the limit of 
negligible interparticle coupling.  Section III then calculates the 
shape of the probability distribution of energy costs for the general 
case of finite nearest neighbor coupling. In Section IV we present 
simulation results for various network geometries.  We also discuss 
the validity of the percolation model and show numerical results for 
the decrease of the threshold with temperature. Section V describes how the current-voltage characteristics behave at temperatures above the point where the voltage threshold reaches zero. Section VI contains a 
discussion of the model and comparisons with recent experimental data 
and as well as with numerical results from related systems.

\section{The basic model in the absence of interparticle capacitive coupling}

We consider one- or two-dimensional arrays of spherical metal 
nanoparticles (``sites''), placed between two in-plane metal 
electrodes.  We ignore any particle-internal level spacing due to 
quantum size effects and treat each site as possessing a continuous 
spectrum of available states up to some local chemical potential. 
This is a reasonable approximation for metal particles with diameters 
larger than a couple nanometers at temperatures above liquid helium. 
For such particles, the largest energy besides thermal energy is the 
electrostatic energy associated with the transfer of additional, 
single electrons.

We consider interparticle spacings small enough to allow for such 
transfer by electron tunneling. We make the usual assumptions of the 
``orthodox theory'' of single electron tunneling (see, e.g., Likharev 
in Ref. \onlinecite{likharev}), namely that the tunnel time is negligible in 
comparison with all other time scales, the tunnel resistance 
$R >> R_q = h/e^2$, where $R_q$ is the quantum of resistance, 
co-tunnel events due to coherent quantum processes can be ignored, and 
the local tunnel rate from site to site depends only on the change in 
electrostatic free energy of the system, $\Delta E$, that would result from a 
tunnel event.  At low temperatures, a positive $\Delta E$ implies a 
suppression of tunneling (Coulomb blockade), and current flows only 
after an external bias has been applied that compensates for this 
energy cost. If tunneling occurs from a site at higher energy to one 
at lower energy ($\Delta E < 0$), we assume that the energy 
difference is lost due to scattering processes in the destination 
particle (inelastic tunneling).

Throughout the paper, we consider the limit of negligible structural 
disorder of the arrays, i.e., all sites are identical in terms of 
both their tunnel coupling and capacitive coupling to neighbors, as 
well as in their self-capacitance.  Disorder enters in form of a 
random distribution of the local chemical potentials at every site 
due to quenched offset charges.  This quenched charge disorder models 
charge fluctuations due to impurities in the substrate which in turn 
polarize the nanoparticles.

A corresponding experimental system can be realized as shown in Ref. \onlinecite{parthasarathy04} by self-assembling, onto an insulating 
substrate, ligand-coated nanoparticles from solution. The ligands 
prevent nanoparticle sintering and well-ordered arrays are formed 
through a balance between attractive van der Waals forces and 
repulsive steric hindrance between ligands from neighboring 
particles.  For dodecanethiol ligands and particle diameters in the 
range 4.5nm to 7nm, a size dispersion of less than 5\% can be 
achieved, resulting in 2D arrays with excellent long-range order of 
the particle packing.  Electronic measurements, both on nanoparticle 
arrays but also on self-assembled monolayers of molecules by 
themselves, have shown that alkanethiol ligands act as mechanical 
spacers and do not otherwise affect the transport properties.\cite{joseph,wang}  Consequently, they set the width of the tunnel 
barrier between neighboring nanoparticles but do not introduce states 
inside the barrier.

The quenched charge disorder is not a perturbative effect: in 
principle, the chemical potential of a nanoscale particle can be 
shifted by a nearby trapped charge as much as it would be by an added 
mobile electron. Therefore, electrons in an array propagate through a 
network of junctions with randomly varying threshold voltages. Note 
that the mobile charges are quantized (electrons) and thus move the 
local chemical potential by the same amount, $\Delta \mu$, every time 
a single charge enters or leaves a site.  On the other hand, the 
quenched charges model a polarization effect and thus can move the 
local chemical potentials continuously, just like a local gate 
electrode could. The overall energy cost, $\Delta E$, associated with a 
tunnel event therefore has to take into account the effect of both 
discrete mobile charges and of a continuous random distribution of 
quenched charges.

One might expect conduction through large arrays to depend on the 
details of the local quenched, or background, charge distribution. 
However, zero-temperature arguments by MW indicate that this is not 
the case,\cite{MW} as least in the limit of negligible capacitive 
coupling between sites.  Instead, the overall array current-voltage 
characteristics (IVs) appear to be robust to background charge 
disorder and exhibit a non-zero effective voltage threshold, $V_t$, that scales linearly with array size (i.e., distance between 
electrodes).

To see this, consider first a 1D array at $T = 0$ with a given 
distribution of quenched polarization charge values. Because mobile 
electrons can compensate for local polarizations in integer multiples 
of $e$, the electronic charge, only disorder in the range $\left[ 
-e/2, +e/2 \right]$ needs to be considered. Starting from an initial 
state of zero applied overall bias, mobile charges can penetrate, say 
from the left, single-file into the disordered potential landscape 
until they first encounter a local up-step in electrostatic 
potential, $\Delta V > 0$.  At this point, the Coulomb blockade 
prevents further advance.

To the left of the up-step, each site now has one additional charge 
on it and all potentials have been raised uniformly by 
$e/C_0$  where $C_0$ is the self-capacitance of each site.  In order 
to move the charge front further toward the right electrode, the bias 
applied to the left electrode has to be raised.  Each time an up-step 
is encountered anywhere in the array, a bias increment of $e/C_0$ at 
the left electrode will suffice to advance the front.  Thus, the 
minimum bias required in order for mobile charges to make it all the 
way across the array will be given by the number of up-steps times 
$e/C_0$ (recall that down-steps in local potential do not 
matter as tunneling is assumed to be inelastic). In other words, the 
$T=0$, global threshold for conduction for an array of $N$ sites is 
given by Ref. \onlinecite{MW} as
\begin{equation}
V_t(0) = \alpha N e/C_0.
\label{mwvt}
\end{equation}
If we now assume a flat, random distribution of quenched charges, on 
average half of the steps between neighboring sites will be up-steps. 
Therefore, for 1D arrays $\alpha = 1/2$.

Note that this argument of MW depends only on the number of up-steps, 
but not on their magnitude $| \Delta V |$! Thus, details of the 
distribution of step sizes are irrelevant at $T=0$.  This also holds 
for 2D systems, except that now the mobile charges can, to some 
extent, avoid up-steps.  Consequently, there will be some roughness 
in the front of charges advancing across the array below threshold. 
Equation \ref{mwvt} still holds, with $N$ now the distance across the 
gap between the electrodes. $N \alpha$ is the number of up-steps in the path across the array with the least number of up-steps (``optimal path''). The value of $\alpha$ in 2D will be smaller than in 1D and depend on the array topology.
Unfortunately, analytical 
arguments that would predict $\alpha$ for 2D systems are not known 
and one has to resort to computer simulations. Specifically, for a 
close-packed triangular arrangement of spheres we find $\alpha = 
0.226$ (see Section IV).

In order to model the effect of finite temperature on the global 
threshold for conduction, we start by considering thermal 
fluctuations at the local, single junction level.  Let $\Delta E$ 
denote the change in the electrostatic potential energy of the system 
when a single electron moves from one site to another. If $| \Delta E 
| >> k_BT$, the nonlinear, Coulomb blockage dominated current-voltage 
characteristic will survive: current will be suppressed below the 
local voltage threshold but will rise approximately linearly above it.\cite{likharev} On the other hand, for $|\Delta E| << k_BT$, the 
Coulomb blockade vanishes and the junction conductance will exhibit 
linear, Ohmic behavior down to the lowest bias voltage.

As a first approximation, we now coarse-grain the system into two 
categories of tunnel junctions. Junctions between sites with energy 
differences $| \Delta E| > bk_BT$ will be treated as if $T = 0$, 
implying a fully nonlinear response and, below threshold, the absence 
of zero-bias conductance.  Junctions between sites with energy 
differences $| \Delta E| < bk_BT$ will be treated as if $\Delta E = 
0$ and all Coulomb blockade effects were removed, implying a linear 
response like Ohmic conductors.  The parameter $b$ measures the 
extent of thermal broadening and depends on details of the electronic 
level distribution. If energy levels are within $bk_BT$, then 
electrons from thermally excited states above the Fermi level on site 
i can tunnel directly into available states below the Fermi level on 
neighboring site j.  This means that up-steps within $bk_BT$ are 
effectively removed.

To determine $b$, we consider in each nanoparticle the width of the 
tail of unoccupied states below and of occupied states above the 
Fermi level. Each tail has an approximate width of $k_BT$ so that $| 
\Delta E|$ is reduced by roughly $2k_BT$ and thus $b \approx 2$.  To 
make this argument more quantitative, we consider the mean energy of 
states above the Fermi energy $\mu$ in particle i,
\[
\langle E_{high} \rangle _i = \frac{\int_{\mu_i}^{\infty} E D(E) f(E) 
dE}{\int_{\mu_i}^{\infty} D(E) f(E) dE}
\]
where $D(E)$ is the density of states and $f(E)$ is the Fermi-Dirac 
function. Evaluating the integral as a series and determining the 
coefficients numerically, we obtain $\langle E_{high} \rangle _i 
\approx \mu_i + 1.2 k_BT$. By symmetry, the mean energy of the 
low-energy unoccupied tail in particle j will be $\langle E_{low} 
\rangle _j \approx \mu_j - 1.2 k_BT$. Tunneling from the high-energy 
tail of particle i to the low-energy tail of particle j thus will 
cost a mean energy $\Delta E = (\mu_j - \mu_i) - 2.4 k_BT$. This leads to 
$b = 2.4$.

As temperature is raised, more and more junctions will satisfy $| 
\Delta E| < bk_BT$ and lose their nonlinear behavior.  We define 
$p(T)$ as the fraction of junctions that has been effectively 
linearized. Since both up- and down-steps will be affected equally by 
thermal smearing, $p(T)$ can be found from
\begin{equation}
p(T) = 2 \int_{0}^{bk_BT} P(\Delta E) d \Delta E
\label{ptint}
\end{equation}
if the distribution of step heights, given by the probability density 
$P(\Delta E)$, is known.  The process of linearizing will happen 
randomly throughout the array until, at some temperature $T^*$, 
sufficiently many junctions have been replaced by Ohmic conductors 
that a continuous path involving only such conductors spans the 
array.  At this point, the overall response will necessarily also be 
linear and the threshold must have reached zero: $V_t(T^*) = 0$.

An upper limit on when this point is reached can be obtained from 
percolation theory by considering the two classes of junctions as two 
types of bonds between neighboring sites.  At small overall bias, we 
can label the nonlinear junctions as insulators and the Ohmic ones as 
conductors.  If a (temperature-dependent) fraction $p(T)$ of all 
junctions in the array has been linearized, and in the absence of 
correlations between neighboring junctions, the first continuous path 
of  linear conductors across the array occurs, on average, at a 
critical fraction $p_c$.  Here $p_c$ is the bond percolation 
threshold which depends only on lattice topology and dimension (for 
corrections due to correlations see Section IV).  Using Eq. 
\ref{ptint}, we thus find $T^*$ through
\begin{equation}
p(T^*) = p_c.
\label{ptst}
\end{equation}

As a consequence of these considerations, the global threshold will 
be a decreasing function of temperature and approach zero as $p 
\rightarrow p_c$.  Hence, to first order,
\begin{equation}
V_t(T) = V_t(0)(1- p(T)/p_c).
\label{vtpt}
\end{equation}

In order to proceed and find the linearized fraction of junctions, 
$p(T)$, we need to know more about the actual distribution $P(\Delta 
E)$ of energy costs.  It will be calculated in detail in Section III. 
However, a few important aspects are already clear from Eq. 
\ref{ptint}.  In particular, since $p_c/2$ is no larger than 1/4 for 
2D lattices,\cite{SE} we have to integrate over only a small 
portion of $P(\Delta E)$ in order to reach a significant suppression 
of the threshold.  If $P(\Delta E)$ does not change much over this 
range, we find
\begin{equation}
p(T) \approx 2bk_BTP(0)
\label{ptpz}
\end{equation}
and $p(T)$ is proportional to temperature. The relevant energy scale,
$1/P(0)$, can be thought of as an effective charging energy, 
while $b$ depends only on the shape of the internal energy 
distribution of the metal particle and thus is independent of 
topology, dimensionality and the effects of coupling.

We will see in Section III that this is a reasonable approximation 
not only for the case of zero capacitive coupling, but even more so 
when nearest neighbor coupling is included.  Physically this is so 
because coupling flattens out the polarization-induced disorder in 
the energy landscape and small energy costs become more probable so 
that $P(\Delta E)$ decays slower for small $\Delta E$.  Combining 
Eqs. \ref{vtpt} and \ref{ptpz} we see that the normalized threshold 
decays linearly with temperature according to
\begin{equation}
\frac{V_t(T)}{V_t(0)} = 1 - 4.8k_BTP(0)/p_c,
\label{vtt}
\end{equation}
where we have used the result $b = 2.4$ obtained earlier. 
In analogy
with the $T=0$ result Eq. \ref{mwvt}, the right hand side of this equation
represents $\alpha(T)/\alpha$, the temperature-dependent number of
up-steps in the optimal path normalized by the number at
zero temperature.

Equation 
\ref{vtt} is a central result of this paper. It predicts a linear 
depression of the global threshold with temperature, with a prefactor 
$2 b k_B P(0) / p_c$ that is universal and does not depend on the details 
of the threshold distribution.

\section{Energy cost distribution including nearest-neighbor coupling}

To calculate $P(\Delta E)$, we start from the electrostatic energy of 
a system of capacitors,
\begin{equation}
E = \frac{1}{2} \sum_{i,j} (q_i + Q_i) {\bf C}^{-1}_{ij} (q_j + Q_j),
\label{esum}
\end{equation}
where the $q_i$ are quenched, offset charges and the $Q_i$ are mobile 
charges (equal to an integer multiple of $e = -1.6 \times 10^{-19}$C 
or zero). The ${\bf C}^{-1}_{ij}$ are elements of the inverse 
capacitance tensor. Note that ${\bf C}^{-1}_{11}$, in the standard 
definition of the capacitance tensor, does include contributions from 
coupling to nearest neighbors if such coupling is present.

We define the energy difference before/after tunneling of a single 
electron from site 1 to site 2 as
\begin{equation}
\Delta E = E_{Q_1=0,Q_2=e} - E_{Q_1=e,Q_2=0}.
\label{deltenon}
\end{equation}
In the absence of any quenched charge disorder ($q_i = 0$) we have 
$\Delta E = 0$, and there is no cost associated with moving charges 
around inside the array. In other words, there is no Coulomb blockade 
of tunneling (even though $\Delta \mu_j > 0$) and the current-voltage 
characteristic will be linear.

Now imagine a flat, random distribution of quenched polarization 
charges in the range $q_i \in \left[ -e/2, +e/2 \right]$.  As before, 
this range suffices because larger offsets will be compensated by 
mobile charges of magnitude $e$. In the limit of negligible 
capacitive coupling between sites considered for now, this leads to
\[
\Delta E = e \left( q_1 - q_2 \right) {\bf C}^{-1}_{11}.
\]

To deal with nearest neighbor capacitive coupling, we focus here on 
the case of a close-packed, triangular lattice simply for the sake of 
having a concrete picture in mind and for direct comparison with 
experiments.  In general, any lattice type can be treated the same 
way and the differences affect only the quantitative results for the 
capacitance tensor elements.

\begin{figure}[tb]
\begin{center}
\includegraphics[width=8.6cm]{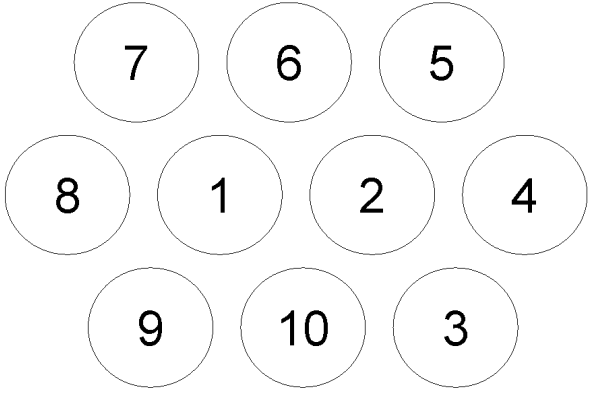}
\end{center}
\caption{Ten-sphere subsystem in a triangular lattice. The electron 
transfer occurs between sites 1 and 2. The other sites are the 
nearest neighbors.
}
\label{spheresg}
\end{figure}

We consider a subset of the triangular lattice consisting of 10 
spheres: two central sites (\#1 and \#2) participating in the 
tunneling event and their 8 surrounding neighbors as in Fig. 
\ref{spheresg}. Keeping only nearest neighbor elements and taking 
$Q_j = 0$ for $j > 2$,
\begin{eqnarray*}
\Delta E & = & e \left( q_1 - q_2 \right) \left( {\bf C}^{-1}_{11} - 
{\bf C}^{-1}_{12}  \right) + \\
& & e {\bf C}^{-1}_{12} \left( q_3 + q_4 + q_5 - q_7 - q_8 -q_9 \right).
\end{eqnarray*}
Defining $\gamma \equiv {\bf C}^{-1}_{12} / {\bf C}^{-1}_{11}$, we 
write $\Delta E$ as
\begin{eqnarray}
\Delta E & = & e^2 {\bf C}^{-1}_{11} \{ \left[ 1-\gamma \right] 
\left( q_1 - q_2 \right) + \nonumber \\
& & \gamma \left( q_3 + q_4 + q_5 - q_7 - q_8 -q_9 \right) \}.
\label{deltesum}
\end{eqnarray}
The terms in round brackets, containing the $q_i$, are sums of 2 or 6 
random variables. The maximum value for $\Delta E$ is achieved if the 
appropriate limiting values ($+e/2$ or $-e/2$) are inserted for the 
$q_i$.  This gives
\[
\Delta E_{max} = e^2 {\bf C}^{-1}_{11} \left(1+2 \gamma \right).
\]

Without capacitive coupling to neighbors, $\Delta E_{max}$ can be 
written as $\Delta E_{max} = e^2/C_0$, where $C_0 = 4 \pi \epsilon 
\epsilon_0 r$, is the capacitance of a single sphere of radius $r$ 
embedded in a medium of dielectric constant $\epsilon$. The 
key points emerging from equations \ref{deltenon} and 
\ref{deltesum} are that the system energy cost associated with a 
tunnel event is not equivalent to
the change in chemical potential of a single 
site, and that existence of a range of polarization charges 
$q_i$ gives rise to a distribution of energy costs $\Delta E$.

To calculate the full distribution $P(\Delta E)$ of energy 
differences, we need to first find the distributions $P_2(x)$ and 
$P_6(x)$ resulting from the addition of 2 or 6 random variables. In 
general, the probability of obtaining a value $x = x_1 + x_2 + ... + 
x_n$ from the sum (or difference) of $n$ independent random numbers 
$x_i$ can be calculated from their 
recursion relation:
\[
P_n(x) = \int_{-\infty}^{\infty} dX^{\prime}
                P_{n-1}(x-x^{\prime}) P_1(x^{\prime}).
\]

Using Fourier transform to convert the convolution into a product,
we get $P_n(\xi)=P_{n-1}(\xi)P_1(\xi)$.
This leads to $P_n(\xi)=P_1^n(\xi)=
\left[\sin(\xi/2)/(\xi/2)\right]^n$, or
\[
P_n(x) = \frac{2}{\pi} \int_{0}^{\infty} \frac{\sin ^n \xi}{\xi ^n} 
\cos \left( 2 \xi x \right) d \xi.
\]
Specifically, for $n=2$ and $n=6$ this integral can be solved 
analytically and gives
\begin{eqnarray*}
P_2(x) & = & (|x-1| + |x+1| - 2|x| ) / 2 \\
P_6(x) & = & (|x-3|^5 + |x+3|^5 - 6 |x-2|^5 - 6 |x+2|^5 +  \\
& & 15 |x-1|^5 + 15 |x+1|^5 - 20 |x|^5 ) / 240
\end{eqnarray*}

\begin{figure}[tb]
\begin{center}
\includegraphics[width=8.6cm]{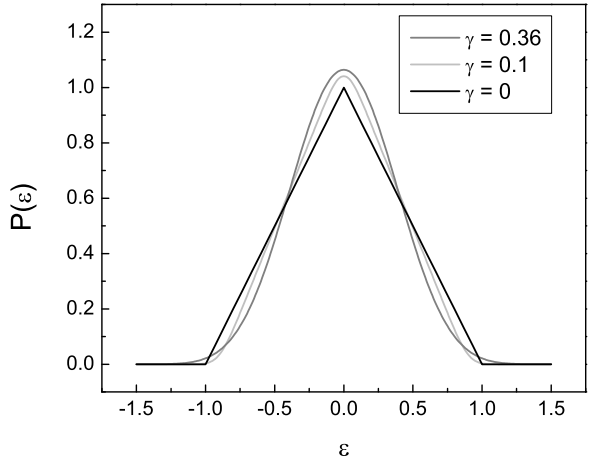}
\end{center}
\caption{
Probability distribution of the energy cost of tunneling between sites 1 and 2 in Fig. \ref{spheresg}. The distribution of $\varepsilon \equiv \Delta E/(e^2{\bf C}^{-1}_{11})$ is plotted where $\Delta E$ is the change in the system energy due to tunneling. ${\bf C}^{-1}_{11}$ is the diagonal element of the inverse capacitance matrix and $\gamma \equiv {\bf C}^{-1}_{12}/{\bf C}^{-1}_{11}$.}
\label{pofdelteg}
\end{figure}

The probability distribution of $\Delta E$ in Eq. \ref{deltesum} is 
then given by
\begin{eqnarray}
P(\Delta E) & = &\frac{1}{e^2 {\bf C}^{-1}_{11}} \int_{- 
\infty}^{+\infty} \frac{1}{\gamma (1-\gamma)} P_2 \left(\frac{\Delta 
E - \Delta E'}{(1-\gamma) e^2 {\bf C}^{-1}_{11}} \right) \times 
\nonumber \\
& & P_6 \left( \frac{\Delta E'}{\gamma e^2 {\bf C}^{-1}_{11}} \right) 
d \Delta E'.
\label{pdelteint}
\end{eqnarray}
The shape of this $P(\Delta E)$ is triangular with apex at $\Delta E 
= 0$. Depending on $\gamma$, the shape is rounded near the top 
(where$\Delta E \rightarrow 0$) and curved outward near the 
bottom (as $\Delta E_{max}$ is approached).  The amount of 
rounding/curving increases with $\gamma$ (Fig. \ref{pofdelteg}). 
Specifically, for negligible coupling ($\gamma=0$),  $P(\Delta E)$ 
becomes the distribution of differences between two random variables
\begin{equation}
P(\Delta E) = 1/\Delta E_{max} - | \Delta E | / \left( \Delta E_{max} \right)^2.
\label{triangle}
\end{equation}
This is a simple triangle with $P(0) = 1/\Delta E_{max}$ and base 
extending from $-\Delta E_{max}$ to $+\Delta E_{max}$.
Fig. \ref{pofdelteg} shows $P(\varepsilon)$ as a function of the 
normalized energy cost, $\varepsilon = \Delta E / (e^2{\bf 
C}^{-1}_{11})$.


Using Eqs. \ref{ptint} and \ref{triangle} for $\gamma = 0$, we find 
that the fraction of linearized junctions is
\begin{equation}
p(T) = \frac{2bk_BT}{\Delta E_{max}} - \left( \frac{bk_BT}{\Delta 
E_{max}} \right) ^2.
\label{pttriangle}
\end{equation}

For a 2D triangular lattice $p_c = 0.347$ so that the temperature at 
which an Ohmic conducting path percolates across the lattice, defined 
in Eq. \ref{ptst} by $p(T^*) = p_c$,  is reached at $bk_BT^*/ \Delta 
E_{max} = 0.192$.  This value is small enough that, to good 
approximation, Eq. \ref{ptpz} holds and the quadratic term in Eq. 
\ref{pttriangle} can be neglected for all $T < T^*$.

For finite capacitive coupling between nearest neighbors, $\gamma > 
0$, $P(\varepsilon)$ in Eq. \ref{pdelteint} can be expanded around 
$\varepsilon = 0$ to obtain
\begin{equation}
P(\varepsilon) = P(0) - \frac{0.55}{\gamma (1-\gamma)^3} 
\varepsilon^2 + {\mathcal{O}}(\varepsilon ^3).
\label{peps}
\end{equation}
The linear term disappears because the distribution has a rounded top 
near $\varepsilon = 0$ (Fig. \ref{pofdelteg}). Consequently, 
corrections to Eq. \ref{ptpz} are of order $(bk_BT^*/\Delta 
E_{max})^3$ and thus smaller than in the case of zero coupling (see 
Section IV for numerical integration results for $T^*$).  Therefore, 
the linear decrease of $V_t$ with temperature in Eq. \ref{vtt} holds 
to even better approximation.
The first term, $P(0)$, in Eq. \ref{peps} can be found 
straightforwardly from the integral in Eq. \ref{pdelteint} as long as 
$\gamma$ is sufficiently small.  This leads to
\[
P(0) = \frac{1}{e^2 {\bf C}^{-1}_{11}} \left[ \frac{1}{1-\gamma} - 
\frac{2\gamma}{\left( 1- \gamma \right)^2} \int^{\infty}_{0} x P_6(x) 
dx \right]
\]
and finally,
\begin{equation}
P(0) \approx \frac{1}{e^2 {\bf C}^{-1}_{11}} \frac{1-1.57 
\gamma}{\left( 1-\gamma  \right)^2}.
\label{pzapprox}
\end{equation}

Note that $P(0)$ depends only on the geometry of the system and is 
independent of all details of the quenched charges (as long as they 
can be assumed uniformly random).  This allows us to obtain $P(0)$ 
from calculations of the capacitance tensor elements ${\bf 
C}^{-1}_{11}$ and ${\bf C}^{-1}_{12}$. In Section IV we present 
numerical results for a range of coupling strengths and show how 
these tensor elements depend on the ratio of center-to-center 
distance, $L$, to particle radius, $r$. As particles get closer and $L/r 
\rightarrow 2.4$, $\gamma$ reaches 0.4 and the approximation leading 
to Eq. \ref{pzapprox} breaks down (see also Fig. \ref{pzecg}a below).
Furthermore, for very large interparticle coupling, next-nearest 
neighbor interactions will become significant and correlations 
between energy-steps may become more important.

We can repeat the above derivation of $P(0)$ for a one-dimensional 
linear chain of particles. In this case, we consider 4 sites in a row 
with an electron moving between the two central sites. Now $P(\Delta 
E)$ contains the integral of a product of two $P_2$ functions. We find 
that for $\gamma < 1/3$,
\begin{equation}
P(0)_{1D} = \frac{1}{e^2 {\bf C}^{-1}_{11}} \frac{1 - 4\gamma/ 3}{(1 
- \gamma)^2}.
\label{pzoned}
\end{equation}

One final aspect concerns how the zero-temperature threshold $V_t(0)$ in
Eq. \ref{mwvt} is affected by capacitive coupling between neighboring
particles.  In MW's argument leading to Eq. \ref{mwvt} for the uncoupled
case, the factor $e/C_0$ came from an increase in local
potential corresponding to one full electronic charge. With capacitive
coupling, the increase in local potential due to an electronic charge
will be less as it effectively spreads out over the neighbors.

In order to reach the threshold for conduction, we  still
have to add approximately one electron to the array for each up-step in a
path. To first order, the average local change in potential associated
with adding an electron is $e {\bf C}^{-1}_{11}$, where
${\bf C}^{-1}_{11}$ decreases with increasing coupling. As before,
$\alpha$ is the number of up-steps in the optimal path at $T=0$ divided by the
length of the array. The optimal path is the one with the fewest number
of up-steps. Let us define $V_0$ as the average increase in external bias required to overcome an up-step. We then can think of the voltage threshold as a product of
two quantities: the number of up-steps ($\alpha N$) and the cost in bias per up-step ($V_0 \approx e {\bf C}^{-1}_{11}$). Modifying Eq. \ref{mwvt}, we are led to
\begin{equation}
V_t(0) = \alpha N V_0 \approx \alpha N e {\bf C}^{-1}_{11}.
\label{vteeff}
\end{equation}

Note, however, that this relation is only an approximation and that a 
full calculation is a formidable problem for $\gamma > 0$. The reason 
is that now local changes in potential depend strongly on 
the quenched charge configuration as well as on other mobile charges 
arriving on nearby particles. In 2D, in particular, this complex 
interaction poses a challenge not only for analytical calculations 
but also for simulations. On the other hand, 1D simulations can be 
carried out straightforwardly and can be used to gauge the validity 
of Eq. \ref{vteeff}. This will be done in the next section.

\section{Numerical calculations and checks }

In order to use Eqs. \ref{vtt} and \ref{vteeff}, we need to know 
certain elements of the inverse capacitance matrix as well as the 
value of $\alpha$ appropriate for a given lattice. Both of these can 
be obtained from numerical calculations as we detail in this Section. 
In addition, simulations allow us to perform a number of checks of 
the assumptions underlying the model developed in Section III and 
they provide a direct test for the effect of correlations that were 
neglected in its derivation. In the following figures, we normalize 
capacitances by the capacitance, $C_0$, of an isolated sphere and 
energies by $e^2/C_0$, the maximum energy cost for tunneling 
between capacitively uncoupled particles.

{\em Inverse Capacitance Matrix.} To calculate the capacitance matrix 
of the 10-sphere system in Fig. \ref{spheresg}, we used 
\textsc{fastcap}, a capacitance extraction program developed at MIT.\cite{fastcap} The program implements a preconditioned, adaptive, 
multipole-accelerated 3D capacitance extraction algorithm developed 
by Nabors {\em et al.}.\cite{nabors}  Each site in the system was 
represented by a spherical, 1200-panel polygon. Center-to-center 
distances between 2.1 and 20 times the radius were examined. (For 
$L/r = 20$, we used a 104-panel sphere approximation so as to not run 
out of computer memory.) The output of the program is a 10x10 
capacitance matrix ${\bf C}$ in units of pF for spheres of radius 1m. We then 
inverted this matrix in Mathematica to find ${\bf C}^{-1}$. Since capacitance is directly proportional to the scale of the system, and to the dielectric constant, we can remove these dependences by scaling all capacitance elements by the self-capacitance of an isolated sphere. We will do this in all the figures to give a general result.

\begin{figure}[tb]
\begin{center}
\includegraphics[width=8.6cm]{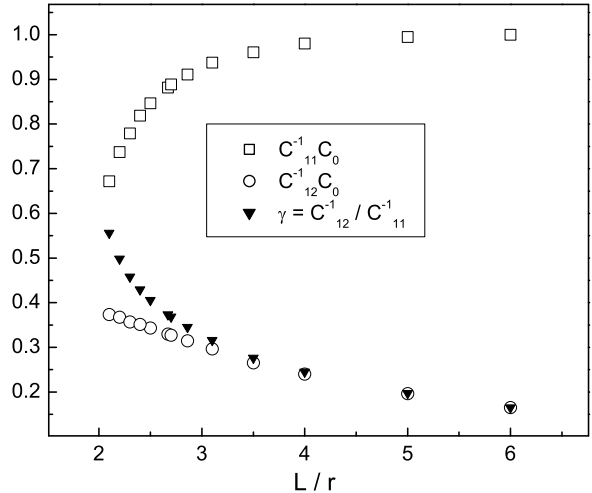}
\end{center}
\caption{
Effect of coupling on the elements of the inverse capacitance matrix for 
a 10-particle triangular system. Coupling increases as the 
center-to-center spacing, $L$, normalized by the radius, $r$, decreases. 
As coupling increases, the inverse self-capacitance, ${\bf 
C}^{-1}_{11}$, decreases and the inverse interparticle capacitance, 
${\bf C}^{-1}_{12}$, increases. We normalize by the 
self-capacitance of an isolated sphere, $C_0$, to assure that the 
values plotted are independent of $r$ and the dielectric constant. 
}
\label{capsg}
\end{figure}

Fig. \ref{capsg} shows the effect of coupling on the 1-1 and 1-2 
elements of the inverse capacitance matrix. Note that the 
self-capacitance ${\bf C}_{11}$ and thus ${\bf C}^{-1}_{11}$ depends 
on interparticle coupling because nearby spheres can polarize when a 
charge is added to the central sphere, decreasing the overall energy 
cost of the charge addition. However, as Fig. \ref{capsg} shows, for 
values of $L/r > 3$ the change in ${\bf C}^{-1}_{11}$ due to 
nearest-neighbor coupling is small, and ${\bf C}^{-1}_{11}$ remains 
within 10\% of $1/C_0$. Typical experimental values for close-packed, 
dodecanethiol-coated 6nm particles give values $L/r$ of about 2.7.\cite{parthasarathy04}  As Fig. \ref{capsg} shows, the off-diagonal 
element ${\bf C}^{-1}_{12}$  depends less strongly on $L/r$ than the 
diagonal element.  Thus, the increase in  $\gamma$ with decreasing 
$L/r$ below a value of about 3 is largely due to ${\bf C}^{-1}_{11}$.

We also note that the interparticle capacitance ${\bf C}_{12}$ 
depends on having extra neighbors.  For example, for $L/r = 2.67$, 
${\bf C}_{12}$ in the 10-sphere system of Fig. \ref{spheresg} is only 
71\% of the value obtained for two isolated spheres. Thus, it is 
essential to look at the system as a whole and not to assume isolated 
spheres. In order to check whether or not the 10-sphere system is sufficient, we added another ring of spheres to Fig. 1, creating a 24-particle subset of the triangular array. We then calculated the full capacitance matrix for the 24-particle system. For $L/r = 2.1$, we found that the changes in ${\bf C}_{11}$ and ${\bf C}_{12}$ are less than 1\%. Consequently, we take the 10-sphere system as a sufficiently good approximation to the triangular array.

The results of the capacitance matrix calculation are in contrast to approximations\cite{Black} which estimate the effect of capacitive coupling by adding to 
the capacitance of an isolated single particle, $C_0$, the
interparticle capacitance $C_{12}$ for each neighbor. In particular, over the 
range $2.1 < L/r < 4$ the estimate ${\bf C}_{11} \approx C_0 + 6 
C_{12}$ for a triangular lattice gives about twice the value for 
${\bf C}_{11}$ obtained numerically using 
\textsc{fastcap}.

In principle, \textsc{fastcap} will give the full 
capacitance matrix of the 10-particle system in Fig. \ref{spheresg}, 
and thus take into account several longer 
range couplings.  However, here we limit the discussion to
nearest neighbor coupling. First, we examine the zero-temperature 
limit.


{\em Conduction Threshold at T = 0.} To calculate numerically the 
onset of conduction at $T = 0$, we follow MW's model and look for 
paths across the array that minimize the number of up-steps. For 
this, we use a variant of the well-known Dijkstra optimal path finding algorithm, the 
``bottleneck algorithm.''\cite{cormen} For each site, we define an 
offset charge $q_i$. If $q_i > q_j$, then i-to-j is considered an energy
up-step in the uncoupled case.  While we cannot use this method to 
find the full current-voltage characteristics, it provides a very 
fast and effective way of determining the validity of Eq. \ref{mwvt} 
and it allows us to extract the geometrical prefactor $\alpha$. As 
defined in Section III, $\alpha$ is the number of up-steps in the 
optimal path at $T=0$ divided by the length of the array. Note that our 
definition of $\alpha$ differs from MW who define $\alpha = V_t(0) 
C_0 / N e$. The two definitions only agree in the uncoupled case.

\begin{figure}[tb]
\begin{center}
\includegraphics[width=2.8cm]{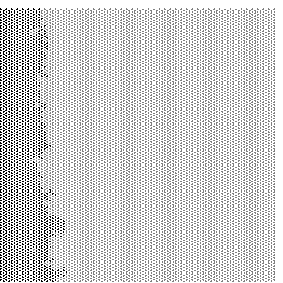}
\hfill
\includegraphics[width=2.8cm]{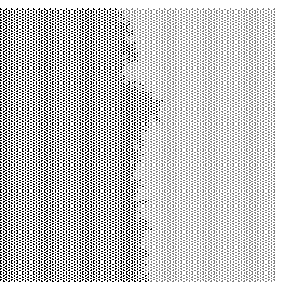}
\hfill
\includegraphics[width=2.8cm]{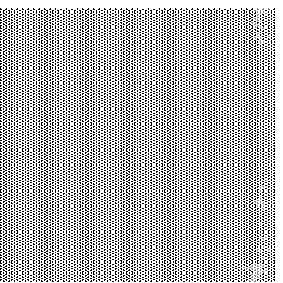}
\end{center}
\caption{Charge front in a 2D triangular lattice as a function of 
external bias. The mobile charges, shown in dark gray, are able to 
penetrate further into the array from a reservoir on the left as bias 
is increased from left to right in the 3 pictures. The simulations on 
a 100x100 array were carried out using the ``bottleneck'' algorithm 
from Ref. 22.
}
\label{thsimulationg}
\end{figure}

We can also numerically obtain the charge front as it propagates 
across the array for voltages below threshold. To do so, we find all 
the sites that can be reached in less than a given number of energy 
up-steps. In Fig. \ref{thsimulationg} we show three snapshots from a 
simulation on a triangular lattice with increasing bias from left to 
right. The advancing charge front is seen as the right-hand edge of 
the dark gray region.


Let us first consider the uncoupled case. For all types of lattices 
investigated, we find that $V_t(0)$ increases linearly with $N$, as 
predicted by Eq. \ref{vteeff}. For a 2D square lattice MW reported 
$\alpha = 0.338(1)$ using Monte Carlo simulations. The bottleneck 
algorithm gives $\alpha = 0.329(7)$ for a 160x160 square lattice 
(averaged over 1000 trials). For honeycomb and triangular arrays 
(100x100 array, 1000 trials) we find $\alpha = 0.301(9)$ and $\alpha 
= 0.226(8)$ respectively.

What is the effect of coupling on the number of up-steps in the 
optimal path? A ``step'' $\Delta E$ between two sites is not just 
$(q_i -  q_j)/C_0$, but now takes into account all neighbors, as in 
Eq. \ref{deltesum}. However, since $\alpha$ does not depend on the 
magnitude of the up-steps, we do not expect a large effect.  This is 
borne out by the simulations. In 1D, $\alpha$ is not affected by 
coupling even for ${\bf C}_{12} / {\bf C}_{11} >> 1$. In a 2D 
triangular array, we find that $\alpha$ depends only weakly on 
coupling. For $L/r = 2.1$, $\alpha$ decreases by about 10\% from its 
uncoupled value; for $L/r \geq 5$, $\alpha$ has essentially the 
uncoupled value of 0.226.

In order to compare our model more directly with literature results 
for the global threshold in the coupled case, which are available 
only in 1D,\cite{MW} we simulated a 1D chain of sites. 
In this simulation, we only consider self-capacitance (${\bf C}^{-1}_{11}$) and nearest-neighbour capacitance (${\bf C}^{-1}_{12}$). An electron moves forward from site i to i + 1 if $\Delta E_{i 
\rightarrow i+1} < 0$, where $\Delta E$ is calculated from Eq. \ref{esum} considering 
both offset charges $q$ and integral charges $Q$ from all previous 
tunneling events on all sites. 

The external bias is raised in 
increments much smaller than $e {\bf C}^{-1}_{11}$ to inject 
electrons into the system. Electrons are allowed to propagate forward 
and rearrange to find the minimum energy state of the system before 
increasing the bias again. $V_t$ is the external bias 
value for which the first electron reaches the far end of the chain. 
For each disorder realization in a 100-site chain, we count the 
number of up-steps and then raise the bias to find the threshold. As 
mentioned in the previous section, for finite coupling, there is no 
unique cost in bias per up-step, but rather a distribution. Fitting 
the average cost, $V_0$, to a quadratic function for $\gamma < 
0.4$, we find
\begin{equation}
V_0 \frac{1}{e{\bf C}^{-1}_{11}} = 1 - 1.93 
\gamma + 1.53 \gamma^2 + {\mathcal{O}}(\gamma^3).
\label{onedsim}
\end{equation}

\begin{figure}[tb]
\begin{center}
\includegraphics[width=8.6cm]{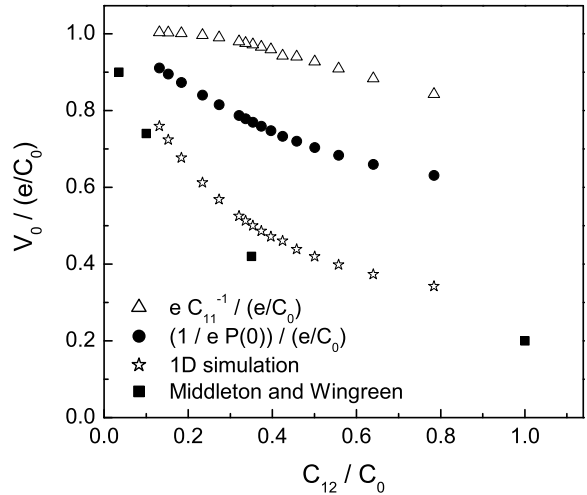}
\end{center}
\caption{
Average  external voltage bias per up-step, $V_0$, at threshold in
a 1D chain of spheres as a function of interparticle coupling at $T=0$. The
vertical axis is normalized by the bias per up-step in the uncoupled
case, $e/C_0$, where $C_0$ is the self-capacitance of an isolated sphere.
The horizontal axis is the interparticle capacitance, $C_{12}$, normalized
by $C_0$. The data from our 1D simulation (open stars) are compared with
simulation results from Ref. 1 (full squares) and two analytical
approximations (open triangles and filled circles).
}
\label{onedg}
\end{figure}

Results from this simulation are shown in Fig. \ref{onedg}, where we 
plot the average cost per up-step, $V_0$, normalized to the uncoupled 
value, as a function of ${\bf C}_{12} / C_0$ in a 1D chain. This is 
compared to the approximations $V_0 \approx e {\bf 
C}^{-1}_{11}$ from Eq. \ref{vteeff} and $V_0 \approx 1 / eP(0)$ 
using the 1D result, Eq. \ref{pzoned} for $P(0)$. Also shown are three 
data points from MW's Fig. \ref{spheresg}, based on a full simulation 
of the current-voltage characteristics of a chain. (Note that MW use a 
different normalization in their Fig. 1, i.e., they plot $V_t C_0 / eN$, and 
extend the simulations to larger coupling strengths.)


\begin{figure}[tb]
\begin{center}
\includegraphics[width=2.8cm]{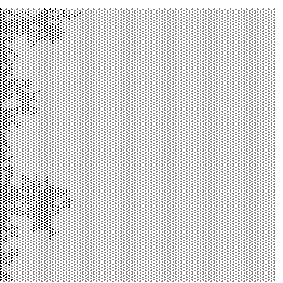}
\hfill
\includegraphics[width=2.8cm]{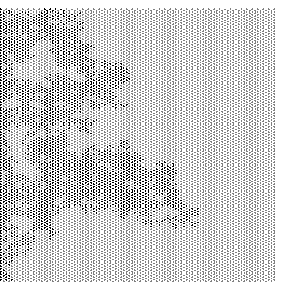}
\hfill
\includegraphics[width=2.8cm]{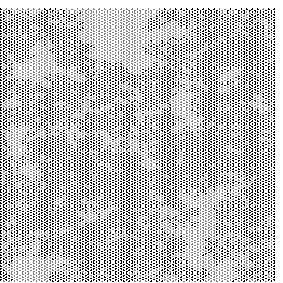}
\end{center}
\caption{
Effect of temperature on an 2D triangular array with quenched 
charge disorder. As temperature is increased from left to right in the 3 
pictures, mobile charges (in dark gray) can penetrate deeper into 
the array without energetic cost. When a percolating dark grey path 
spans the array from left to right, the global threshold bias for conduction 
reaches zero.
The simulations on a 100x100 triangular array were carried out using the 
``bottleneck'' algorithm from Ref. 22.
}
\label{tsimulationg}
\end{figure}

{\em Conduction Threshold for T $>$ 0.} As a next step, we add 
temperature to the simulations. In the 2D algorithm that finds the 
optimal path across the array we have direct access to all bonds, and 
thus energy costs, $\Delta E$, for moving an electron between any pair 
of neighboring sites.  This allows us to test the validity of the 
linear decrease of the threshold with increasing temperature predicted by Eq. \ref{vtt}. As temperature increases, 
steps with magnitudes smaller than a threshold energy, $\Delta E_{th} 
= bk_BT$ are thermally erased. In the path-finding algorithm, we count
all steps less than $\Delta E_{th}$ as ``down-steps'', that is, they do not cost any energy. 
We then traverse the energy landscape to find the least-cost path as 
before. Fig. \ref{tsimulationg} shows three snapshots from the 
simulation. The threshold $\Delta E_{th}$, and thus temperature, is 
increased from left to right. In dark grey we show all sites 
reachable without cost from the left edge.


\begin{figure}[tb]
\begin{center}
\includegraphics[width=8.6cm]{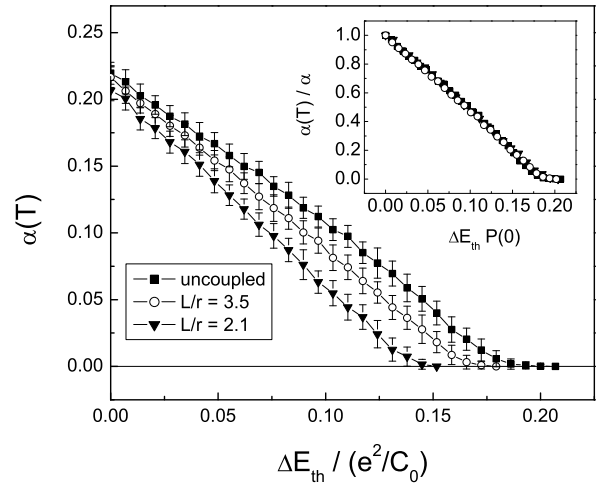}
\end{center}
\caption{
Decrease of the number of energy up-steps in the optimal path with
temperature. $\alpha(T)$ is
defined as the temperature-dependent number of up-steps in the least cost
path across the array divided by the array length. The
effect of thermal fluctuations is introduced by counting up-steps only if
they exceeded a cut-off energy $\Delta E_{th} = bk_BT$. Data are shown from simulations on a 2D triangular lattice
containing 100x100 spherical particles for three
different coupling strengths, parameterized by the ratio of
center-to-center distance, $L$, to sphere radius, $r$. The inset shows the collapse of the curves upon normalization by $\alpha \equiv \alpha(T=0)$, and by $1/P(0)$, the relevant energy scale for temperature-dependence.
}
\label{vttsimulationg}
\end{figure}

In Fig. \ref{vttsimulationg} we plot $\alpha(T)$ as a function of the 
effective temperature, $\Delta E_{th}$, for various degrees of 
coupling. In all cases, we find that the number of up-steps in the 
optimal path decreases with increasing threshold energy approximately 
linearly. (The deviations from a strictly linear decrease, close to 
$\alpha(T) = 0$, come from a finite size effect: the simulated arrays 
contained 100x100 sites, so around $\alpha(T) = 0.01$, the average 
number of up-steps reaches 1, below which the average is fractional and thus no longer a physical measure.) We 
also see that for a given ``temperature'' the threshold decreases 
with increasing coupling. Furthermore, the temperature $T^*$ at which 
$\alpha(T) = 0$ decreases with increasing coupling (see also Fig. 
\ref{pzecg}b). In accordance with the results in Fig. \ref{capsg}, 
these trends are most pronounced for small $L/r$ and saturate near 
the uncoupled behavior for $L/r > 5$.


{\em Percolation and Correlations.} In the analytic calculation of 
$T^*$ in Section II, we found the fraction of linearized bonds, 
$p(T)$, through Eq. \ref{ptint} and defined $T^*$ as $p(T^*) = p_c$, 
where $p_c$ is the bond percolation threshold in the lattice under 
consideration.  This procedure relies on two assumptions that we now 
test.

First, the basic idea of our tunneling model is that none of the 
down-steps cost energy.  Implicit in Eq. \ref{ptint} is a somewhat 
more restrictive criterion, namely $|\Delta E| < bk_BT$, requiring that 
the path be along only those bonds (corresponding to either up- or 
down-steps) that had been linearized by thermal fluctuations. There 
may be energetically much more optimal, but more asymmetric, paths 
that take advantage of those larger-energy down-steps that have not 
yet been linearized.  In this situation, we are starting with a 
lattice with all down-steps in place and ask when the system-spanning 
path forms in the process of adding up-steps of increasing size.

Second, setting $p(T^*) = p_c$ and using literature values for $p_c$ 
assumes that the usual rules for bond percolation apply and, in 
particular, all bonds are placed completely randomly.  However, while 
the site energies are from a flat distribution, the energy 
differences between sites are correlated.  For example, in the 
triangular lattice in Fig. \ref{spheresg}, the energy differences 
between sites 1 and 2 and between sites 2 and 6 completely specify 
$\Delta E$ between sites 1 and 6. Therefore, $p_c$ may not 
necessarily provide an accurate value for the threshold. Finally, 
even in the absence of correlations, the finite size of arrays 
corresponding to experimental situations (with $N$ no more than a few 
100) may lead to a small correction to $p_c$ as listed for infinite 
lattices.

We investigated these questions for a variety of lattice types by 
calculating numerically the threshold $p_a$, defined as the average 
fraction of bonds required for percolation under the {\em asymmetric} 
condition $\Delta E < bk_BT^*$, and the threshold $p_s$, defined as 
the average fraction of bonds required for percolation under the {\em 
symmetric} condition $| \Delta E| < bk_BT^*$.  Both are listed in 
Table 1, together with $p_c$ as obtained from the same lattice but 
with randomly assigned bond energies rather than random site 
energies. z is the coordination number, the number of nearest neighbors of each site. 
The lattice with z = 2 consists of 200 parallel 1D wires.

\begin{table}
\begin{ruledtabular}
\begin{tabular}{|l|l|l|l|l|}
z & $p_{c,th}$  & $p_c$ & $p_s$ & $p_a$ \\ \hline\hline
2 & 1 & 0.970(7) & 0.968(39) & 0.945(45) \\ \hline
3 & 0.653 & 0.643(9) & 0.627(15) & 0.539(15) \\ \hline
4 & 0.5 & 0.500(8) & 0.509(12) & 0.433(11) \\ \hline
5 & & 0.420(6) & 0.455(10) & 0.384(10) \\ \hline
6 & 0.347 & 0.346(6) & 0.397(8) & 0.335(9) \\ \hline
7 & & 0.292(6) & 0.348(8) & 0.289(7) \\ \hline
8 & & 0.250(5) & 0.307(7) & 0.255(7) \\
\end{tabular}
\caption{
Percolation coefficients for different coordination numbers 
z, calculated for 200x200 arrays and averaged over 200 trials. 
$p_s$ is the average fraction of bonds that need 
to be linearized in the whole array such that the first 
system-spanning path appears containing only linearized bonds (both 
up- and down-steps). $p_a$ is the average fraction of bonds 
linearized in the array for the first system-spanning path containing 
non-linearized down-steps as long as all up-steps are linearized. $p_s$ and $p_a$ are for systems with random site energies. $p_c$ is the bond percolation fraction for the uncorrelated bond percolation.
The theoretical values 
$p_{c,th}$ are taken from Ref. 19 and presented for comparison to indicate the extent of finite-size effects.
}
\end{ruledtabular}
\end{table}

\begin{figure}[tb]
\begin{center}
\includegraphics[width=8.6cm]{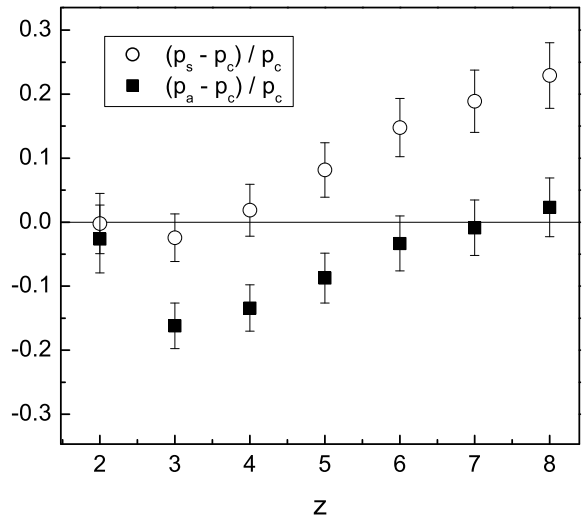}
\end{center}
\caption{
Comparison of symmetric and asymmetric percolation conditions for 
various 2D arrays with coordination numbers, z. 
$p_s$, $p_a$ and $p_c$ are defined in Table 1.
All simulation data are for arrays of size 200x200 and averaged over 200 disorder realizations.
}
\label{pcg}
\end{figure}

\begin{figure}[tb]
\begin{center}
\includegraphics[width=8.6cm]{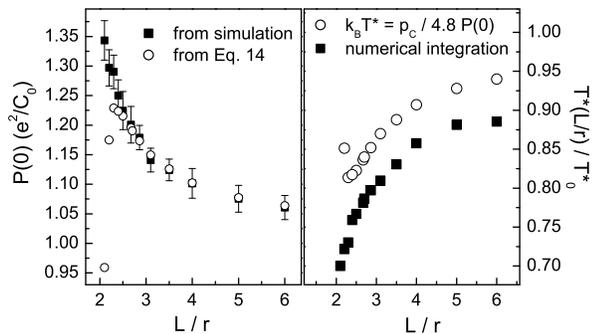}
\end{center}
\caption{
Effect of capacitive coupling on the energy scales associated 
with the temperature-dependence of the global threshold for 
conduction. $L$ is the center-to-center distance between neighboring 
spherical particles and $r$ is their radius. a) Changes in the peak 
of the energy-cost distribution, $P(0)$, as a function of coupling. 
$1/P(0)$ plays the role of an effective charging energy; in this plot 
it has been normalized by the maximum energy cost in the uncoupled 
case, $e^2/C_0$. Results from Monte Carlo data (full 
squares) and from the approximation given by Eq. \ref{pzapprox} (open 
circles) are shown.  b) Changes in $T^*$, the temperature at which 
the voltage threshold of the array becomes zero, for a triangular 
lattice (percolation threshold, $p_c = 0.347$). $T^*$ 
has been normalized by $T^*_0$, the value in the uncoupled case. 
Closed square symbols are data from numerical integration of the 
energy cost distribution, $P(\Delta E)$, as obtained by Monte Carlo 
simulation; open circles show the approximation given by Eq. 
\ref{ptpz}.
}
\label{pzecg}
\end{figure}

In Fig. \ref{pcg}, a comparison between $p_s$ and $p_c$ gives a sense 
of the relevance of correlations which increase $p_s$ roughly 
linearly with increasing z (ignoring the case of z = 2).  At the same 
time, antisymmetric paths involving large down-steps give a 
threshold fraction, $p_a$, that is systematically lower than $p_s$ by 
about 15\%.  Intriguingly, and quite unexpectedly, for lattices with 
z = 6 to z = 8 the contributions from correlations and asymmetry 
appear to cancel each other to a large extent so that $p_c$ provides an 
excellent estimate of the ``true'' value, $p_a$. Thus, using $p_c$ in 
Eq. \ref{ptst} to estimate $T^*$ should give very reasonable 
estimates for experiments on self-assembled nanoparticle layers. The 
small difference between the theoretical $p_c$ and the value from the 
simulation shows the insignificance of finite size effects for 200x200 arrays.


{\em Distribution of Energy Costs.} The last approximation in the 
model we wish to test is the replacement of the integral in Eq. 
\ref{ptint} with $p_c = 4.8 k_BT^* P(0)$.
To find the full distribution of energy costs for the nearest 
neighbor-coupled 10-sphere system shown in Fig. \ref{spheresg}, we 
used a Monte Carlo routine.  Offset charges from a uniform random 
distribution $\left[ -e/2, +e/2 \right]$ were assigned to each of the 
10 sites. Using the capacitance matrix as calculated from 
\textsc{fastcap}, the energy cost $\Delta E$ associated with 
tunneling from site 1 to site 2 was found from Eq. \ref{deltesum} for 
each disorder realization. $P(\Delta E)$ was then obtained 
from sampling $\Delta E$ for $10^6$ offset charge realizations for 
each value of $L/r$.

Fig. \ref{pzecg}a shows the normalized peak probability density $P(0)e^2/C_0$ 
as a function of $L/r$. $P(0)e^2/C_0$ only depends on $L/r$ since both $C_0$ and $1/P(0)$ are 
proportional to $r$ and $\epsilon$. We compare the simulation value with the approximation in Eq. \ref{pzapprox}. Knowing the full distribution from Monte Carlo simulations allows us to find, without 
approximations, the critical temperature $T^*$, where the voltage 
threshold goes to zero.  According to Eqs. \ref{ptint} and \ref{ptst} 
this is done by integrating $P(\Delta E)$ out to the point where the 
area under the graph corresponds to $p_c$. In Fig. \ref{pzecg}b, we 
compare the results of numerical integration with the analytical 
approximation $T^* = p_c / ( 2bk_BP(0) )$ (Eq. \ref{ptpz}) with $P(0)$ from Eq. \ref{pzapprox} for a 2D triangular lattice.


\section{Current-Voltage Characteristics above $T^*$}

Next, we investigate the behavior for temperatures $T \simeq  T^*$ and
above.  Within our model, $T^*$ is defined as the temperature at which
there are just enough local junctions linearized to span the array at
zero bias and remove the global threshold.  In other words, with
increasing temperature the nonlinear current-voltage ($I-V$)
characteristics, described by the powerlaw $I \sim (V - V_t(T))^{\zeta}$,
have been linearly shifted to the left until, at $T^*$, they first reach
the origin with finite slope. This gives rise to a finite zero-bias
conductance, $g_0 \equiv dI/dV|_{V=0}$. 

For $T > T^*$, additional linearized
junctions provide parallel paths across the array and the zero-bias
conductance increases. For any given temperature, however, an increase in
bias will eventually provide sufficiently high local voltage drops to
involve portions of the array with junctions not yet linearized by
thermal fluctuations.  Thus, at sufficiently high bias, the $I-V$
characteristics will change back from Ohmic to the original nonlinear
powerlaw behavior with temperature-independent exponent $\zeta$.

These considerations correspond to a picture in which the nonlinear $I-V$
curve of a 2D array emerges from summing the contributions from all paths
carrying current at a given bias voltage (this is the essence of the
scaling derived by MW). As the bias is increased above $V_t(T)$, more and
more paths will open up as their up-steps are overcome. Above $T^*$, when
the global threshold disappears along the optimal path, there still are
many other paths that have finite thresholds and are accessible at higher
bias. These thresholds will keep decreasing linearly with temperature as
more and more steps are linearized by thermal fluctuations. As a
consequence, the high-current, powerlaw portion of the $I-V$ curves will
shift to lower and lower bias voltages.

We therefore expect the $I-V$ curves to collapse, by a simple horizontal
shift, onto a master curve not only for $T < T^*$, but, at least with
their high-bias portion, also for $T > T^*$. In the latter regime a true
threshold for global conduction obviously no longer exists and
$V_t(T)$, as obtained from the shift required for a high-bias $I-V$
collapse onto a common power law, should be thought of as an {\em
effective} threshold value.  This effective $V_t(T)$ will be negative for
$T>T^*$.

These predictions are borne out by experiments.  As first observed by
Ancona {\em et al.}\cite{ancona} and also seen in Fig. \ref{expg}, the linear
decrease of $V_t$ with $T$ continues past $T^*$ and into a regime of
negative effective threshold values.  (Our simulation results in Fig. \ref{vttsimulationg}
are not based on calculating full $I-V$ curves. Since our method finds
the first system-spanning path, it is no longer applicable above $T^*$
where such paths exist already at zero applied bias.) 

We note that our
model provides a natural explanation for this smooth cross-over from
positive to negative $V_t(T)$ which does not invoke additional
mechanisms. In particular, it does not bring into play activation over
the tunnel barriers for $T>T^*$, as proposed by Ancona {\em et al.}.\cite{ancona}
Because such barriers are set by the properties of the alkanethiol
ligands separating adjacent nanoparticles, we expect barrier heights in
excess of several eV, corresponding to the first available electronic
states in the ligands.  This is significantly higher than either $k_BT$
or the voltage drop per up-step, making hopping over the barrier highly
unlikely.

How does $g_0$ depend on temperature?  To address this point, we revisit
a simplifying approximation made in the model, namely the coarse-graining
in which non-linearized junctions were assumed to be unaffected by
thermal fluctuations and to exhibit zero conductance below threshold. In
principle, of course, finite $T$ will always induce some zero-bias
conductance. For a single junction this zero-bias conductance exhibits
activated behavior, i.e., is proportional to $\exp(-U/k_BT)$, where the
activation energy, $U = \Delta E$, is the energy cost required to move a
charge across the junction.\cite{likharev}

Sufficiently far below $T^*$,
there always will be several junctions along the optimal path with
$\Delta E >> k_BT$.  Consequently, the overall zero-bias conductance will
be exponentially suppressed to a level where $g_0$ is well approximated
by the coarse-graining approximation.

Once $T^*$ is approached, linearized
junctions for the first time form a system-spanning path.  $g_0$ will be dominated by the relatively few bottleneck
junctions with the largest activation energy in the path, $\Delta E  \sim k_BT^*$. The majority of paths around
the bottlenecks would involve junctions with much larger $\Delta E$ which
therefore could shunt the bottlenecks only insignificantly.\cite{bottleneck1,bottleneck2}

Therefore, near $T^*$ the overall,
zero-bias array conductance, $g_0$, will display activated behavior similar
to a single junction with $U = bk_BT^* = p_c / (2P(0))$. As before, the
key point here is that the activation energy is not simply the charging
energy of an isolated grain, but is connected to the optimal path across
an energy landscape established by the quenched charge disorder. For 2D
triangular arrays $p_c/2 \sim 0.17$ so that $U$ is approximately 1/5 of
the effective charging energy $1/P(0)$.

Above $T^*$, additional paths in which all up-steps have been thermally
erased will span the array.  These parallel paths will contribute to
$g_0$ and modify the behavior. Taking the number density, $D(T)$, of such
paths to be proportional to the percolation conductivity of the
linearized subset of junctions above $p_c$, we have $D(T) \sim (p(T) -
p_c)^t$, where $p(T)$ is given by Eq. \ref{ptint} and $t \approx 1.3$ in 2D.\cite{SE} The overall zero-bias conductance at temperature $T$
then follows from integration of $D(T')g_0(T')$ between $T^*$ and $T$.
This will give a powerlaw correction to the simple activated behavior,
but we expect the optimal path established at $T^*$ to continue to
dominate since its bottlenecks have the lowest activation energy of the
set.

The experimental data in Ref. \onlinecite{parthasarathy04}  for the zero-bias
conductance in 2D arrays above $T^*$ is compatible with simple activated
behavior, with values for $U$ that are within a factor two of $p_c /
(2P(0))$. However, since $T^*$ can reach 100K or more in these arrays, the
remaining interval up to room temperature simply is not large enough to
provide a stringent test of the model predictions. The simple activated
form for $g_0$ was also observed by Black {\em et al.} in
arrays of Co nanoparticles.\cite{Black}

\section{Discussion}

The model described in the preceding sections provides a physical 
picture for the role of thermal fluctuations as they affect the 
nonlinear transport properties in systems with random local 
thresholds for conduction. One key aspect is that such systems cannot 
be described by associating a single, fixed energy cost with charge 
motion from site to site. Instead, it is important to consider the 
fact that energy costs depend on both the Fermi levels of the 
particles involved in the tunneling process, and on the 
surrounding charge environment. This is the essence of Eq. \ref{deltenon}, which 
gives the change in electrostatic energy of the system as a whole, 
and, in the nearest neighbor approximation considered here, leads to 
the probability density of energy costs shown in Fig. \ref{pofdelteg}. 

In 
particular, the energetics of the system are not determined simply by 
the change in the Fermi level (by some fixed single electron 
``charging energy") of the nanoparticle tunneled into.  Energy costs 
derive from differences in the total system energy before and after a 
tunneling event and thus involve at a minimum two particles or, with 
capacitive nearest neighbor coupling, ten particles in a 2D 
triangular lattice (Fig. \ref{spheresg}) or four particles along a 1D line. In the 
absence of quenched charge disorder, these energy costs vanish and 
mobile charges can move freely inside the array (the only costs are 
incurred when charges enter the edges of the array from one of the 
electrodes). Therefore, the scaling of the global threshold with 
system size $N$ in Eq. \ref{mwvt} (the original MW result) or its modification 
for finite coupling strength, Eq. \ref{vteeff}, are a direct consequence of 
quenched charge disorder.

However, as far as the global threshold is concerned, the detailed 
shape of the full energy cost distribution turns out to be not 
critical.  Rather, as we showed by mapping the system onto an 
equivalent percolation problem, the behavior is dominated by the 
lowest-cost percentiles (up to $p_c/2$).  To very good approximation, 
this is captured by the zero-cost, peak value of the probability 
density, $P(0)$ (Eqs. \ref{pzapprox} and \ref{pzoned}, for 2D and 1D systems, 
respectively). Increased capacitive coupling is found to have two 
effects:  it flattens the energy landscape, thereby narrowing the 
width of the energy cost distribution and increasing the value of 
$P(0)$, and it rounds off the peak of the distribution, making $P(0)$ 
an even better approximation of the relevant portion of $P(\Delta E)$.

There are several predictions that emerge from the analytic model.

- First, the model predicts a linear decrease of the overall, global 
threshold with temperature (Eq. \ref{vtt}). This decrease is directly 
proportional to $k_BT P(0)$, the strength of thermal fluctuations 
measured relative to the effective charging energy $1/P(0)$. What is 
particularly appealing about this result is that all details about 
capacitive coupling and particle geometry enter through $P(0)$, while 
the numerical prefactor, $4.8/p_c$, captures the underlying network 
topology and dimension through the percolation threshold, $p_c$. 
Therefore, data from systems with similar structure but different 
particle sizes, spacings, or dielectric constants should collapse 
onto a ``universal" curve when plotted as $V_t(T)/V_t(0)$ versus $k_BT 
P(0)$.

- Second, for sufficiently broad distributions of quenched charge 
disorder we expect that thermal fluctuations do not alter the basic 
character and roughness of the energy landscape.  Thus we expect MW's 
zero-temperature powerlaw scaling of the nonlinear current-voltage 
($I-V$) characteristics, $I \sim (V - V_t(0))^{\zeta}$, to survive at 
finite $T$ with fixed exponent $\zeta$, but with $V_t(0)$ replaced by 
$V_t(T)$. In other words, the shape of the $I-V$ characteristics 
remains unaffected by temperature while they are shifted linearly 
towards smaller threshold values.

- Third, the threshold is expected to vanish and the nonlinear, 
Coulomb-Blockade-type current voltage characteristics are expected to 
change to linear, Ohmic behavior near zero bias once the temperature 
exceeds $T^* = p_c/(4.8k_B P(0))$ (Eqs. \ref{ptst} and \ref{ptpz}).

- Fourth, also for finite capacitive coupling, the zero-temperature 
global threshold value, $V_t(0)$, can be written as a product of 
two quantities: the average number of
up-steps encountered along the optimal path across the array, $\alpha N$,
which depends mainly on array geometry, and the average applied voltage
per up-step, $V_0$, at threshold. We argued 
that, at least to first order, capacitive coupling can be taken into 
account by $V_0 \approx e{\bf C}^{-1}_{11}$ (Eq. \ref{vteeff}) instead of $e/C_0$ for the uncoupled case (Eq. \ref{mwvt}).

The robustness of the linear decrease of the global threshold for 
conduction with temperature is underscored by recent simulations of 
the full current-voltage characteristics for mobile charges hopping 
between traps in a 2D lattice.\cite{reichhardt03b} The charges 
interact via a long-range Coulomb term and the trap depth at each 
site is chosen from a Gaussian distribution. In this work Reichhardt 
and Olson Reichhardt not only find behavior qualitatively similar to 
our model, but also track as a function of temperature the charge 
flow patterns beyond threshold. Their results support a basic, 
underlying tenet of our approach, namely that small thermal 
fluctuations simply shift $V_t(T)$ to smaller values while preserving 
the roughness of the energy landscape and thus the shape of the 
current-voltage characteristics.  

Conversely, such linear shift might 
be taken as an indicator of a sufficiently wide distribution of 
trapping depths or local threshold values: simulations by Rendell {\em et al.} of small 2D tunnel junction networks 
using the full, temperature-dependent ``orthodox" Coulomb blockade 
model show this shape-preserving shift energing once quenched charge 
disorder is introduced.\cite{rendell}

\begin{figure}[tb]
\begin{center}
\includegraphics[width=8.6cm]{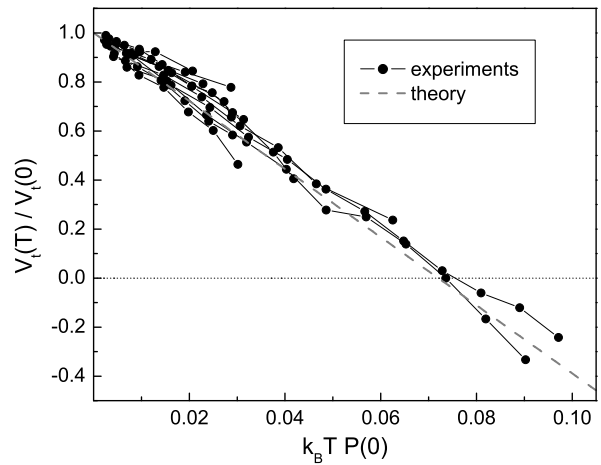}
\end{center}
\caption{
Decrease of the normalized voltage threshold as a function of 
effective temperature variable $k_BTP(0)$ for 6nm diameter, 
close-packed gold nanoparticle monolayers. (Reproduced from Ref. 12 
with permission.)
}
\label{expg}
\end{figure}

The fact that $P(0) e^2/C_0$, through ${\bf C}^{-1}_{11}$ and ${\bf 
C}^{-1}_{12}$, depends only on $L/r$ makes comparison with experiments 
straightforward as long as the array geometry is known.  In 
particular, if $L$ and $r$ can be determined from transmission 
electron micrographs, Fig. \ref{pzecg} together with $C_0 = 4 \pi \epsilon 
\epsilon_0 r$ give direct access to $P(0)$ for triangular arrays, and 
thus make it possible to plot the normalized threshold, 
$V_t(T)/V_t(0)$ as a function of scaled temperature, $k_BT P(0)$. 
Fig. \ref{expg} shows such plot for temperature-dependent threshold 
data obtained from self-assembled, close-packed gold nanocrystal 
monolayers covering a range of array lengths ($27 < N < 170$) and 
effective charging energies $96{\rm meV} < 1/P(0) < 302{\rm meV}$.\cite{parthasarathy04}  The threshold values were obtained from 
linear shifts of the full $I-V$ curves onto a single powerlaw 
mastercurve for each sample (with temperature-independent  exponent 
$ \zeta = 2.25 \pm 0.1$). 


All of these data are seen to cluster 
around the linear decay with slope $-4.8/p_c = -13.8$ and x-intercept 
$k_BT^*P(0) = p_c/4.8 = 0.07$ predicted by the model. Note that this 
data collapse contains no free parameters once $V_t(T)$ has been 
measured (over a wide enough range to extrapolate reliably to 
$V_t(0)$) and $P(0)$ been obtained from the array geometry.  If 
direct access to $L$ and $r$ is not possible, but $N$ can be 
estimated from the electrode spacing, $P(0)$ can be estimated using 
Eq. \ref{vteeff} together with the appropriate value for $\alpha$ listed in 
Section IV.

The linear suppression of $V_t(T)$ with temperature was also observed 
by Ancona {\em et al.}\cite{ancona} in experiments on 2D arrays of 
gold nanoparticles, and by Bezryadin, Westervelt and Tinkham\cite{bezryadin} in studies of 1D carbon nanoparticle chains. While 
micrographs allowing for a determination of $L$ and $r$ were not 
available from either experiment, Bezryadin {\em et al.} found that 
the voltage threshold decreased as $V_t(T) \approx V_t(0) - N 
k_BT/e$. The authors also give the radius of the carbon particles as well as $N$ and $V_t(0)$. From this information we estimate a cost per up-step of about $0.2 e/C_0$. From Fig. \ref{onedg} and capacitance calculations of a 4-particle 1D chain, we find $P(0) = 1.66(C_0/e^2)$. 
With $p_c = 1$ for a 1D chain our model gives $V_t(T) = V_t(0)- 4.8 V_t(0)P(0) k_BT $ from Eq. \ref{vtt}. The second term can be written as $-4.8 [0.5 N (0.2 e/C_0)] (1.66 C_0/e^2) k_BT$, where the term in square brackets is $V_t(0)$ and $\alpha_{1D} = 0.5$.\cite{MW} Using the experimental parameters given by Bezryadin {\em et al.}, our theory predicts $V_t(T) = V_t(0) - 0.8 N k_BT/e$, which is close to the experimental temperature-dependence.

\section{Conclusions}

Our model describes the effect of temperature on the global threshold for
conduction through arrays with distributed local thresholds due to
quenched charge disorder. It resolves two long-standing issues, namely
how to extend the original, $T=0$ scaling approach by MW to $T>0$ and how
to include capacitive coupling between nearest neighbors.  

One key
finding is a robust, linear decrease of the global threshold with
temperature, in excellent agreement with recent experiments on
close-packed nanocrystal arrays. This explains the experimental finding
that powerlaw $I-V$ characteristics resulting from Coulomb blockade
effects keep their nonlinear shape to remarkably high temperatures while
simply being parallel-shifted as $T$ is increased.  The model further
predicts the existence of a cross-over temperature $T^*$, above which the
low-voltage portion of the $I-V$ characteristics changes and acquires a
significant zero-bias conductance that exhibits simple activated
behavior. 

A second key finding is the identification of $1/P(0)$ as the
relevant, effective charging energy, extending earlier results that did
not treat capacitive coupling.  Our approach explicitly takes into
account the fact that quenched charge disorder leads to a distribution of
energy costs for tunneling, even for otherwise perfect lattices of
identical junctions. Finally, we present numerical calculations that
allow one to extract the relevant capacitance matrix elements as well as
$P(0)$ from knowledge of interparticle spacing and particle diameter. 

\begin{acknowledgments}
We would like to thank Raghu Parthasarathy, Alan Middleton, Ilya 
Gruzberg, Philippe Guyot-Sionnest, Tom Rosenbaum and Tom Witten for 
insight and helpful discussions. This work was supported by the 
UC-ANL Consortium
for Nanoscience Research and by the NSF MRSEC program under DMR-0213745.
\end{acknowledgments}

\end {document}